\documentclass[%
 reprint,
superscriptaddress,
nofootinbib,
 amsmath,amssymb,
 aps,
prl,
floatfix,
]{revtex4-2}

\usepackage{graphicx}
\usepackage{dcolumn}
\usepackage{bm}
\usepackage{multirow}

\usepackage{aas_macros}
\usepackage{xspace}
\usepackage{color}
\usepackage{hyperref}

\newcommand{\eg}{\mbox{e.\,g.,}\xspace}

\newcommand{\ie}{\mbox{i.\,e.,}\xspace}

\definecolor{darkgreen}{rgb}{0,0.5,0}
\definecolor{mypink1}{rgb}{0.858, 0.188, 0.478}
\definecolor{red}{rgb}{1,0,0}

\begin{document}

\preprint{APS/123-QED}
\title{$5 \sigma$ tension between \textit{Planck} cosmic microwave background and eBOSS Lyman-alpha forest and constraints on physics beyond $\Lambda$CDM}

\author{Keir K. Rogers}
 \email{k.rogers24@imperial.ac.uk}

\affiliation{Department of Physics, Imperial College London, Blackett Laboratory, Prince Consort Road, London, SW7 2AZ, United Kingdom}

\affiliation{%
Dunlap Institute for Astronomy and Astrophysics, University of Toronto, 50 St.\,George Street, Toronto, ON M5S 3H4, Canada}%

\author{Vivian Poulin}
 \email{vivian.poulin@umontpellier.fr}
 
\affiliation{Laboratoire univers et particules de Montpellier (LUPM), Centre national de la recherche scientifique (CNRS) et Universit\'e de Montpellier, Place Eug\`ene Bataillon, 34095 Montpellier C\'edex 05, France}

\date{\today}

\begin{abstract}
We find that combined \textit{Planck} cosmic microwave background, baryon acoustic oscillations and supernovae data analyzed under $\Lambda$CDM are in 4.9$\sigma$ tension with eBOSS Ly$\alpha$ forest in inference of the linear matter power spectrum at wavenumber $\sim 1 h\,\mathrm{Mpc}^{-1}$ and redshift = 3. Model extensions can alleviate this tension: running in the tilt of the primordial power spectrum ($\alpha_\mathrm{s} \sim -0.01$); a fraction $\sim (1 - 5)\%$ of ultra-light axion dark matter (DM) with particle mass $\sim 10^{-25}$ eV or warm DM with mass $\sim 10$ eV. The new DESI survey, coupled with high-accuracy modeling, will help distinguish the source of this discrepancy.
\end{abstract}

\maketitle

\textbf{Introduction} -- The Standard Models of particle physics and cosmology have been experimentally confirmed with high precision \citep{Planck:2018vyg,Workman:2022ynf}. Yet, open questions remain including the origin of the masses of neutrinos, the source of density fluctuations that evolved into the cosmic large-scale structure (LSS) and the fundamental natures of dark matter (DM) and dark energy (DE).

High-precision datasets such as cosmic microwave background (CMB) temperature T, polarization E,B and lensing \(\phi\) anisotropies \citep{Planck:2019nip,ACT:2023kun,PhysRevD.108.023510}, baryon acoustic oscillations \citep[BAO;][]{Bautista:2020ahg} and supernovae magnitudes \citep[SNe;][]{Brout:2022vxf} are consistent with the standard \(\Lambda\) cold dark matter (\(\Lambda\)CDM) cosmological model. However, there is disagreement with some other datasets in the inference of cosmological parameters, which is either a sign of unmodeled systematics or deviation from \(\Lambda\)CDM. A local distance ladder using SNe and cepheid stars infers the expansion rate today \(H_0\) to be \(\sim 5 \sigma\) higher than from an inverse distance ladder using CMB, BAO and SNe data in the \(\Lambda\)CDM model \citep{Riess:2021jrx}. Further, some LSS data, in particular the clustering and lensing of photometric galaxy surveys like the Dark Energy Survey \citep[DES;][]{DES:2021wwk} and the Kilo-Degree Survey \citep[KiDS;][]{Heymans:2020gsg}, show a \(2 - 3 \sigma\) preference, compared to CMB, BAO and SNe data, for a lower value of the cosmological parameter \(S_8 \equiv \sqrt{\frac{\Omega_\mathrm{m}}{0.3}} \sigma_8\), where \(\Omega_\mathrm{m}\) is the total matter density and \(\sigma_8\) is the amplitude of density fluctuations averaged over \(8 h^{-1}\,\mathrm{Mpc}\) scales today.

\begin{figure}
\includegraphics[width=\columnwidth]{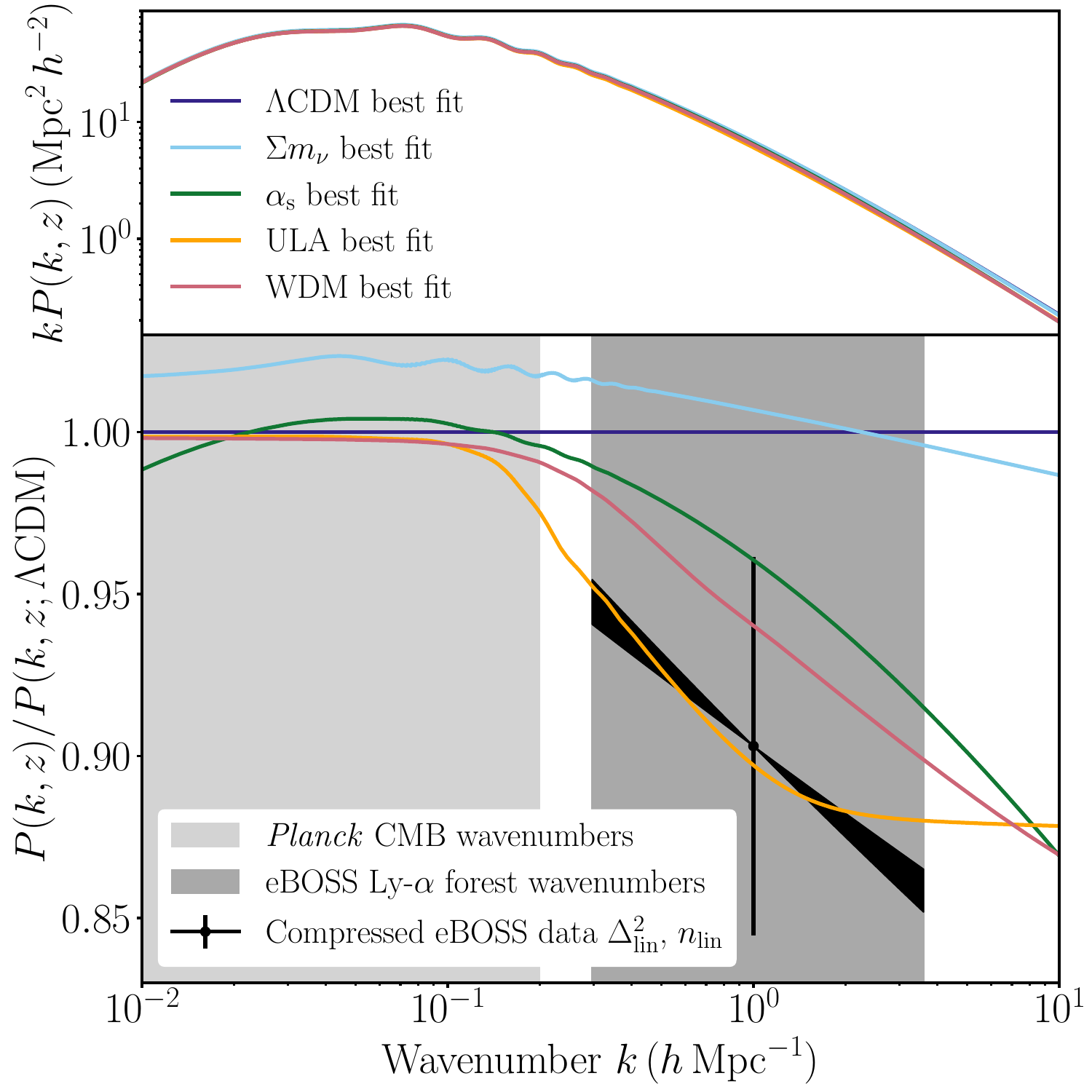}
\caption{\label{fig:power}Best fit linear matter power spectra at \(z = 3\) given \textit{Planck} CMB, BAO, SNe and eBOSS Lyman-alpha forest data for the indicated cosmological models. In the bottom panel, we show the ratio to the best fit \(\Lambda\)CDM model. The vertical black errorbar indicates the \(1 \sigma\) uncertainty on the amplitude \(\Delta_\mathrm{lin}^2\) of the linear matter power at \(z_\mathrm{p} = 3\), \(k_\mathrm{p} = 0.009\,\mathrm{s}\,\mathrm{km}^{-1} \approx 1 h\,\mathrm{Mpc}^{-1}\) as constrained by eBOSS; the diagonal black band indicates the \(1 \sigma\) uncertainty on the tilt \(n_\mathrm{lin}\). The grey bands show the approximate range of wavenumbers to which \textit{Planck} CMB and eBOSS Lyman-alpha forest are sensitive.}
\end{figure}

CMB data are currently sensitive to the matter power spectrum for wavenumbers \(k \lesssim 0.2 h\,\mathrm{Mpc}^{-1}\) \citep[Fig.~\ref{fig:power}; see also][]{ACT:2023kun} through the generation of lensing anisotropies. A key probe of physics beyond \(\Lambda\)CDM is therefore the high-\(k\) (small scale) matter power spectrum. Indeed, the \(S_8\) discrepancy may be a hint of a small-scale deviation from the standard model as galaxy clustering and lensing data from DES and KiDS are modeled for \(k \lesssim 1 h\,\mathrm{Mpc}^{-1}\). Resolutions to the \(S_8\) discrepancy center around scale-dependent changes to the growth of perturbations, \eg ultra-light axion (ULA) DM \citep{Rogers:2023ezo}, time-dependent changes, \eg DM-DE interactions \citep{Murgia:2016ccp,Lucca:2021dxo,Poulin:2022sgp}, decaying DM \citep{Enqvist:2015ara,Pandey:2019plg,FrancoAbellan:2020xnr,FrancoAbellan:2021sxk,Simon:2022ftd}, or stronger galactic feedback \citep{Amon:2022azi}. However, interpretation is complicated by uncertainties like non-linearities, the intrinsic alignment of galaxy shapes \cite{Arico:2023ocu} and a lack of clarity as to the scales at which a discrepancy between CMB and LSS occurs (see below).

The Lyman-alpha forest, quasar spectral absorption formed in the intergalactic medium (IGM), is a powerful probe of high-\(k\), quasi-linear modes for \(k \gtrsim 0.3 h\,\mathrm{Mpc}^{-1}\) \citep[Fig.~\ref{fig:power}; see also][]{Chabanier:2019eai}. This probe therefore provides an independent test of consistency between CMB and LSS data uncomplicated by the problems outlined above, albeit with its own uncertainties regarding the astrophysical state of the IGM. This power arises because it is sourced by moderate IGM fluctuations (tracing underlying DM fluctuations with overdensities \(\lesssim 10\)) that are much less affected by non-linearities and galactic feedback than equivalent scales in galaxy surveys \citep[\eg][]{Lukic:2014gqa}.

The extended Baryon Oscillation Spectroscopic Survey (eBOSS) completed observation of \(\sim 200,000\) quasar spectra with detectable Lyman-alpha forest \citep{Lyke:2020tag}. The high-\(k\) power spectrum can be probed by measuring correlations only along the line of sight. The eBOSS Collaboration reports a 1D Lyman-alpha forest flux power spectrum for velocity wavenumbers \(0.001\,\mathrm{s}\,\mathrm{km}^{-1} < k_\mathrm{F} < 0.02\,\mathrm{s}\,\mathrm{km}^{-1}\) (approximately equating to the spatial wavenumbers shaded in Fig.~\ref{fig:power}) for redshifts \(2.1 < z < 4.7\) \citep{Chabanier:2018rga}. Analyzing the data with a suite of IGM hydrodynamical simulations \citep{Borde:2014xsa}, they constrain \(\Lambda\)CDM parameters \(\sigma_8, \Omega_\mathrm{m}\) and the tilt of the primordial power spectrum \(n_\mathrm{s}\). They find no significant cosmological parameter tension with CMB data \citep{Chabanier:2018rga,Palanque-Delabrouille:2019iyz}.

In this work, we show that the apparent agreement between datasets arises from considering only standard \(\Lambda\)CDM parameters. Using a compressed likelihood that captures all the information contained in the data for the cosmological models that we consider, we assess consistency at the scales and redshifts to which eBOSS data are sensitive, measuring a \(\sim 5 \sigma\) discrepancy with CMB, BAO and SNe data. We then carry out the first analysis of the eBOSS flux power spectrum for a number of extended cosmological models finding that they alleviate the tension.

\textbf{Data likelihoods} -- For a number of cosmological models (\(\Lambda\)CDM, massive neutrinos, power spectrum running, spatial curvature), all the cosmological information in the eBOSS flux power spectrum can be compressed to the amplitude \(\left(\Delta_\mathrm{lin}^2 \equiv \frac{k_\mathrm{p}^3 P_\mathrm{lin}(k_\mathrm{p}, z_\mathrm{p})}{2 \pi^2}\right)\) and tilt \(\left(n_\mathrm{lin} \equiv \frac{\mathrm{d} \mathrm{ln} P_\mathrm{lin} (k, z)}{\mathrm{d} \mathrm{ln} k} |_{k_\mathrm{p}, z_\mathrm{p}}\right)\) of the linear matter power spectrum \(P_\mathrm{lin} (k, z)\) evaluated at a central redshift \(z_\mathrm{p} = 3\) and wavenumber \(k_\mathrm{p} = 0.009\,\mathrm{s}\,\mathrm{km}^{-1}\). This compression holds for models that are sufficiently close to Einstein-de Sitter expansion for the redshifts probed by data and where the linear matter power spectrum is well approximated by a power-law relation for the wavenumbers probed by data \citep{Pedersen:2020kaw,Pedersen:2019ieb,Pedersen:2022anu,McDonald_2000}. Further, when combining with external CMB data, it is accurate to use the marginal likelihood on \(\Delta_\mathrm{lin}^2, n_\mathrm{lin}\) \citep{Pedersen:2022anu}. In the Supplemental Material, we confirm that the compression holds for mixed ULA and WDM models. We therefore use a compressed 2D Gaussian likelihood derived from the eBOSS (Sloan Digital Sky Survey Data Release 14; SDSS DR14) Lyman-alpha forest flux power spectrum \citep{Chabanier:2018rga} and tabulated in Ref.~\cite{Goldstein:2023gnw}\footnote{The likelihood of Ref.~\cite{Chabanier:2018rga} is not publicly available and so Ref.~\cite{Goldstein:2023gnw} tabulated it (in their Table I) by fitting a 2D Gaussian to a \(30 \times 30\) grid of likelihood samples from Fig.~20 of Ref.~\cite{Chabanier:2018rga}. There is a slight non-Gaussian feature that is not captured but, if the full likelihood were available, we do not anticipate that our conclusions would change. Further, Ref.~\cite{Palanque-Delabrouille:2019iyz} extends the analysis of Ref.~\cite{Chabanier:2018rga} by testing extended cosmological models, but the data and likelihood are the same in both analyses; the astrophysical priors are different in the two analyses.}: the mean values of \(\Delta_\mathrm{lin}^2\) and \(n_\mathrm{lin}\) are respectively \(0.310\) and \(-2.340\), the standard deviations are respectively 0.020 and 0.006 and the correlation coefficient is 0.512 (Fig.~\ref{fig:contours_tension}). We give details in the Supplemental Material on the astrophysical model which is marginalized. We also sometimes use a compendium of \textit{Planck} 2018 CMB \citep{Planck:2018vyg,Planck:2019nip}, galaxy BAO \citep{BOSS:2016wmc,Beutler:2011hx,Ross:2014qpa} and Pantheon+ SNe \citep{Brout:2022vxf} data that constitutes a larger-scale complement to the smaller scales probed by the eBOSS flux power spectrum; full details are given in the Supplemental Material.
\textbf{Models and inference} -- We consider four extensions to the baseline six-parameter $\Lambda$CDM model: (i) varying the sum of neutrino masses $\Sigma m_\nu$; (ii) the running of the primordial spectral index \(\alpha_\mathrm{s} \equiv \frac{\mathrm{d}n_\mathrm{s}}{\mathrm{d ln}k}\); (iii) ultra-light axions with particle mass $m_{\rm ULA}$ forming a fraction \(f_\mathrm{ULA}\) of the DM; (iv) warm dark matter with particle mass $m_{\rm WDM}$ and temperature $T_{\rm WDM}$ (forming a fraction \(f_\mathrm{WDM}\) of the DM). The linear matter power spectrum and other cosmological background observables are calculated for these models using the Einstein-Boltzmann solver \texttt{AxiCLASS} \citep{Poulin:2018dzj,Smith:2019ihp}. We use a uniform prior on $\Sigma m_\nu, \alpha_\mathrm{s}, \mathrm{log} [m_{\rm ULA}], f_{\rm ULA}, \mathrm{log} [m_{\rm WDM}], T_{\rm WDM}$ and the standard \(\Lambda\)CDM parameters. We restrict the ULA and WDM fractions to be no greater than 0.05 to ensure that we satisfy the condition for using the compressed likelihood while still optimizing the joint fit to CMB and Lyman-alpha forest data. More details are given on the models, prior and posterior sampling method in the Supplemental Material. The neutrino and running models are constrained by eBOSS Lyman-alpha forest data in Ref.~\cite{Palanque-Delabrouille:2019iyz} to which we compare our results below.


\begin{table}
\begin{ruledtabular}
\begin{tabular}{lccc}
Model & Tension (\(\sigma\)) & \(\Delta \chi^2\) & Parameter constraints\\
\hline
\(\Lambda\)CDM & \(4.90\) & -- & --\\
\(\Sigma m_\nu\) & \(4.80\) & -1.9 & \(\Sigma m_\nu < 0.110\,\mathrm{eV}\)\\
\(\alpha_\mathrm{s} \equiv \frac{\mathrm{d}n_\mathrm{s}}{\mathrm{d ln}k}\) & \(0.92\) & -25.61 & \(\alpha_\mathrm{s} = -0.0108 \pm{0.0022}\)\\
\multirow{2}{*}{ULA} & \multirow{2}{*}{\(0.56\)} & \multirow{2}{*}{-27.58} & \(\mathrm{log} [m_\mathrm{ULA} (\mathrm{eV})] = -24.9 \pm^{1.5}_{1.1}\)\\
& & & \(f_\mathrm{ULA} = 0.0146 \pm^{0.0014}_{0.0086}\)\\
\multirow{2}{*}{WDM} & \multirow{2}{*}{\(1.34\)} & \multirow{2}{*}{-27.08} & \(\mathrm{log} [m_\mathrm{WDM} (\mathrm{eV})] = 1.01 \pm^{0.30}_{0.44}\)\\
& & & \(f_\mathrm{WDM} = 0.0219 \pm^{0.0030}_{0.0042}\)\\
\end{tabular}
\end{ruledtabular}
\caption{\label{tab:bounds} \textit{From left to right}, the parameter tension under the model indicated in the left-most column between \textit{Planck} CMB + BAO + SNe and eBOSS Lyman-alpha forest data; the change in the chi-square \(\Delta \chi^2\) for the combined dataset relative to \(\Lambda\)CDM; and 1D marginalized posterior mean and 68\% credible limits on extended model parameters (95\% upper limits on \(\Sigma m_\nu\) and \(m_\mathrm{WDM}\)). Although, for completeness, we quote marginalized summaries for all extended models, for the neutrino model, there remains tension and so constraints are unreliable. Further, the ULA and WDM posteriors are highly non-Gaussian (Fig.~\ref{fig:contours_constraints}) so summary statistics must be understood in that context.}
\end{table}

\begin{figure}
\includegraphics[width=\columnwidth]{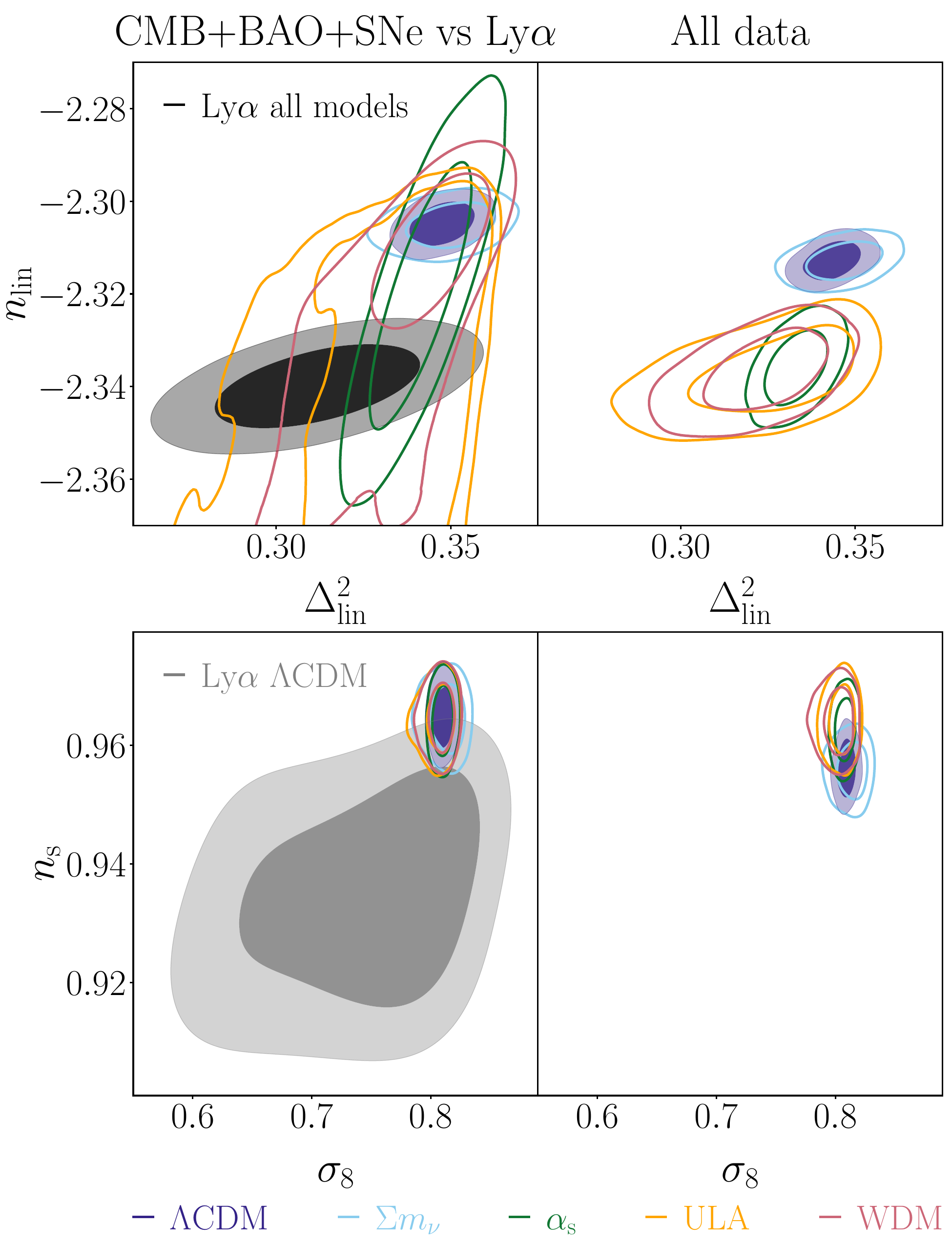}
\caption{\label{fig:contours_tension}Marginalized posterior distributions of \(\Delta_\mathrm{lin}^2, n_\mathrm{lin}, \sigma_8, n_\mathrm{s}\) given \textit{Planck} CMB + BAO + SNe (left panels, colored contours), eBOSS Lyman-alpha forest only (Ly\(\alpha\); left panels, black and gray contours) and the combination of all data (right panels) and the cosmological models indicated by the corresponding color. We show the 68\% and 95\% credible regions. For the top left panel, the Ly\(\alpha\) forest constraint is the same for all models.}
\end{figure}


\textbf{Results} -- Figure \ref{fig:contours_tension} shows the marginalized posterior of \(\Delta_\mathrm{lin}^2, n_\mathrm{lin}, \sigma_8, n_\mathrm{s}\). For the models that we consider, all the cosmological information in the eBOSS Lyman-alpha forest is contained in \(\Delta_\mathrm{lin}^2, n_\mathrm{lin}\) and so the posterior of those parameters is the same for each model. When the information in the flux power spectrum is projected to \(\sigma_8, n_\mathrm{s}\), which characterize the amplitude and tilt of the linear matter power spectrum at different redshifts and smaller wavenumbers, the relative constraint on these parameters is weaker than for \(\Delta_\mathrm{lin}^2, n_\mathrm{lin}\). As a consequence, there is broad agreement between CMB + BAO + SNe and Lyman-alpha forest data when considering only those parameters even under the \(\Lambda\)CDM model. In the Supplemental Material, we explore the use of prior information, in particular on \(H_0\) in order to set the conversion from native velocity to spatial wavenumbers, but still find broad consistency with CMB + BAO + SNe. However, when projecting the information in \textit{Planck}, BAO and SNe data to the larger pivot wavenumber that eBOSS probes, there is a \(4.90 \sigma\) tension\footnote{We measure tension in the non-Gaussian marginalized \(\Delta_\mathrm{lin}^2, n_\mathrm{lin}\) constraints using the parameter difference posterior distribution as implemented in the \texttt{tensiometer} package \citep{Raveri:2021wfz}.} under \(\Lambda\)CDM (Table \ref{tab:bounds}) driven by a suppressed small-scale power spectrum amplitude and tilt. This tension is not alleviated when varying neutrino masses since the suppression is step-like at the wavenumbers to which eBOSS is sensitive (Fig.~\ref{fig:power}). However, the tension is reduced to \(0.92 \sigma\), \(1.34 \sigma\) and \(0.56 \sigma\) respectively for the one-parameter power spectrum running and two-parameter WDM and ULA models; these models suppress both power amplitude and spectral index.

For completeness, we show joint posteriors under all the models we consider (right panels of Fig.~\ref{fig:contours_tension} with parameter constraints and $\Delta \chi^2$ given in Table \ref{tab:bounds} and the Supplemental Material); however, the \(\Lambda\)CDM and neutrino constraints are unreliable as the CMB and Lyman-alpha forest data remain in \(\sim 5 \sigma\) tension under these models. The running, ULA and WDM models are able to recover lower values of \(n_\mathrm{lin}, \Delta_\mathrm{lin}^2\) that eBOSS favors, while maintaining \(n_\mathrm{s},\sigma_8\) consistency with \textit{Planck}. The physics driving these results is elucidated by the minimum-likelihood (best fit) linear matter power spectra in Fig.~\ref{fig:power}. Increasing the mass of neutrinos causes a step-like suppression relative to \(\Lambda\)CDM; in attempting to fit the small-scale tilt, the neutrino masses are set too low to cause sufficient suppression. The best improvements in the total \(\chi^2\) arise from the running, ULA and WDM models.

\begin{figure}
\includegraphics[width=0.49\columnwidth]{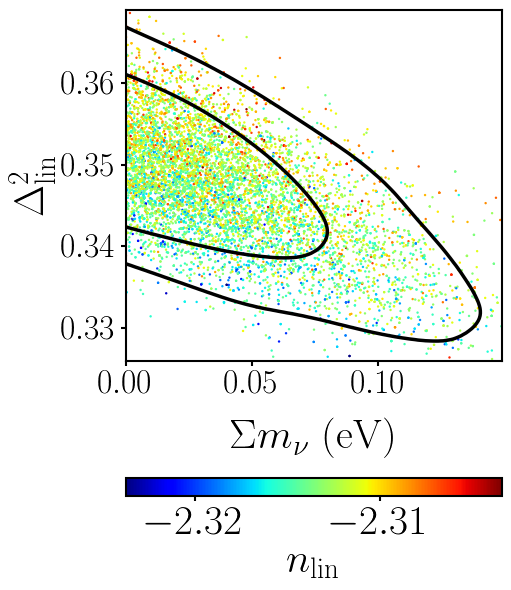}
\includegraphics[width=0.49\columnwidth]{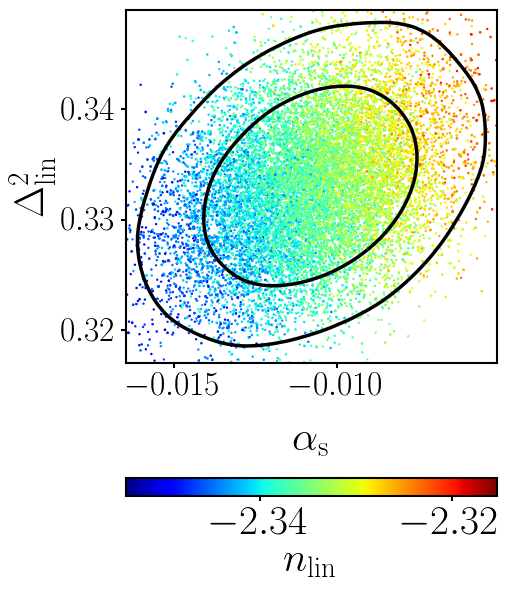}
\includegraphics[width=0.49\columnwidth]{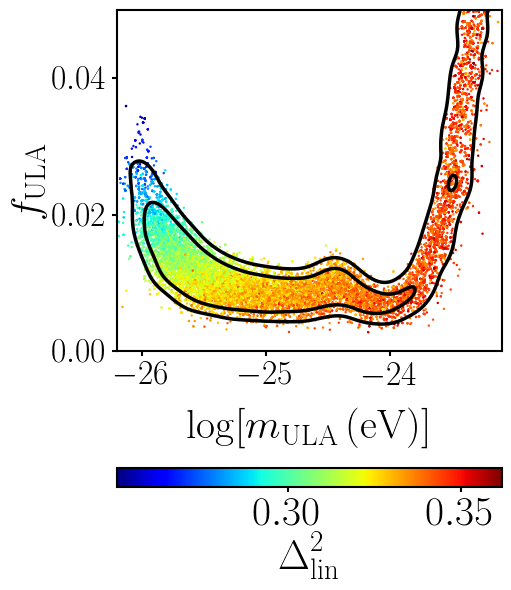}
\includegraphics[width=0.49\columnwidth]{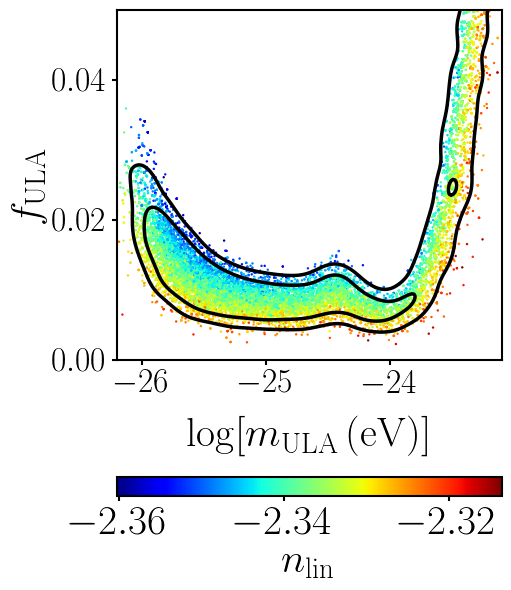}
\includegraphics[width=0.49\columnwidth]{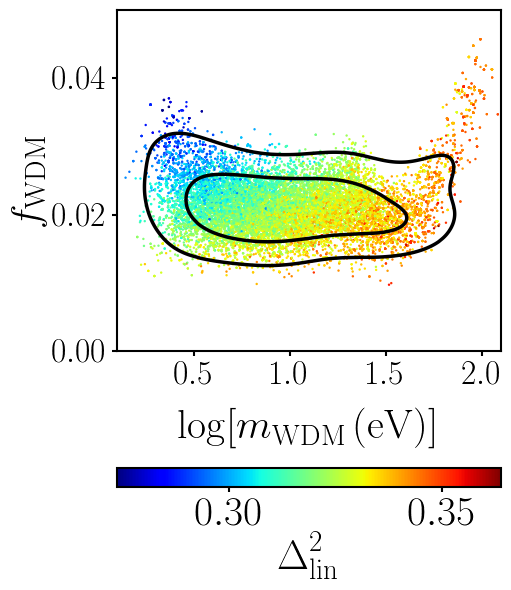}
\includegraphics[width=0.49\columnwidth]{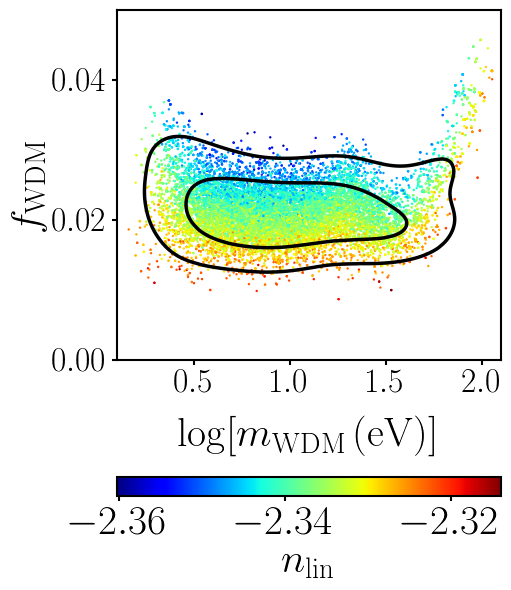}
\caption{\label{fig:contours_constraints}Marginalized posterior distributions of extended cosmological model parameters and \(\Delta_\mathrm{lin}^2, n_\mathrm{lin}\) given \textit{Planck} CMB + BAO + SNe + eBOSS Lyman-alpha forest. We show the 2D 68\% and 95\% credible regions and indicate dependence on a third parameter by the color.}
\end{figure}
Fig.~\ref{fig:contours_constraints} shows joint posteriors of extended model parameters (full sets of 1D and 2D marginalized posteriors in the Supplemental Material). Increasing neutrino masses marginally suppresses the small-scale amplitude but cannot lower the small-scale tilt sufficiently. Increasing the deviation from no running can successfully lower the small-scale amplitude and tilt in agreement with Ref.~\cite{Palanque-Delabrouille:2019iyz} (see Supplemental Material for results where we simultaneously vary \(\Sigma m_\nu\) and \(\alpha_\mathrm{s}\)). The ULA constraint, which disfavors the case with no axions, has a characteristic ``w'' shape. The WDM constraint shows an anti-correlation between particle mass and temperature with marginalized limits $\mathrm{log} [m_{\rm WDM}\,(\mathrm{eV})] = 1.01\pm^{0.30}_{0.44}$ and $\frac{T_{\rm WDM}}{T_{\rm CMB}} = 0.194\pm^{0.043}_{0.074}$, where \(T_\mathrm{CMB}\) is the CMB temperature, giving $f_{\rm WDM} = 0.0219\pm^{0.0030}_{0.0042}$ (all mean and 68\% c.l.). For both ULA and WDM, as the particle mass increases, the suppression wavenumber increases until it is larger than the eBOSS pivot and so doesn't lower \(\Delta_\mathrm{lin}^2\). For a given mass, there is a range of ULA fractions and WDM temperatures that moves the power spectrum transfer function to have the correct amplitude and tilt.

\textbf{Discussion and conclusions} -- We show that, while eBOSS and \textit{Planck} show broad consistency in inference on \(\sigma_8, n_\mathrm{s}\), they are in \(4.9 \sigma\) tension in \(\Delta_\mathrm{lin}^2, n_\mathrm{lin}\) under the \(\Lambda\)CDM model. This result arises since eBOSS probes wavenumbers (\(k \sim 1\,h\,\mathrm{Mpc}^{-1}\)) larger than those to which \(\sigma_8\) (\(k \sim 0.25\,h\,\mathrm{Mpc}^{-1}\)) and \(n_\mathrm{s}\) (\(k \sim 0.07\,h\,\mathrm{Mpc}^{-1}\)) are sensitive, while also unable, without external information, to constrain simultaneously all \(\Lambda\)CDM parameters. When the eBOSS constraint on \(\Delta_\mathrm{lin}^2, n_\mathrm{lin}\) is projected to \(\sigma_8, n_\mathrm{s}\), it weakens; whereas, the CMB + BAO + SNe constraint on \(\Lambda\)CDM is sufficiently strong that it cannot accommodate the suppressed small-scale matter power spectrum amplitude and tilt that eBOSS measures. We conclude that it is crucial to assess consistency between datasets at the wavenumbers and redshifts to which data are sensitive; otherwise, inconsistency can be missed.

We find that varying the masses of neutrinos cannot resolve this parameter discrepancy. This means that constraints on neutrino masses (when not varying any additional parameters) from \textit{Planck} CMB and eBOSS Lyman-alpha forest are derived from datasets in significant tension and may not therefore be reliable; Ref.~\cite{Palanque-Delabrouille:2019iyz} presents limits when varying neutrino and running parameters (see also Supplemental Material)\footnote{Indeed, this inconsistency applies for any model that does not induce suppression and running on small scales.}. We set the first constraints on the ULA and WDM fractions from eBOSS Lyman-alpha forest; we find a preference for ULAs with \(m_\mathrm{ULA} \sim 10^{-25}\,\mathrm{eV}\) and \(f_\mathrm{ULA} \sim (1 - 5) \%\) improving the best fit by \(\Delta \chi^2 = -27.6\) over \(\Lambda\)CDM with no residual tension ($0.56\sigma$). Similarly, we find a contribution of WDM improves the fit by \(\Delta \chi^2 = -27.1\) with \(m_{\rm WDM} \sim 10\) eV, \(T_{\rm WDM} \simeq 0.19\,T_{\rm CMB}\) and reduced tension ($1.34 \sigma$). With one fewer free parameter, running of the power spectrum  reduces the tension to \(0.92 \sigma\) and improves the fit by \(\Delta \chi^2 = -25.6\) \citep[see also][]{Palanque-Delabrouille:2019iyz}.

A key effect that changes the tilt of the matter power spectrum inferred from the Lyman-alpha forest is residual contamination from high-column-density (HCD) absorbers \citep{Borde:2014xsa,Chabanier:2018rga,Palanque-Delabrouille:2019iyz}. The eBOSS Collaboration found that using a more accurate model for residual HCDs \citep{Rogers:2017bmq,Rogers:2017eji}, calibrated on \textit{Illustris} hydrodynamical simulations \citep{Vogelsberger:2014kha,Bird:2014pia}, does not improve the flux power spectrum fit. However, there are other systematic effects related to HCDs that are ignored in 1D Lyman-alpha forest analyses, but which our results suggest should now be investigated: \eg the masking of HCDs in data is correlated to the large-scale structure; the efficiency of detection and masking is uncertain and so the residual population is unknown. Further, Ref.~\cite{10.1093/mnras/staa1242} demonstrated that mis-modeling of baryonic feedback from active galactic nuclei (AGN) can bias inference on \(n_\mathrm{s}\) by \(\sim 2 \sigma\) using the Horizon-AGN simulations \citep{10.1093/mnras/stw2265}. Given that the \(S_8\) discrepancy can be explained by extremely strong feedback models \citep{Amon:2022azi}, coupled with our new results, it is timely to consider the effect of feedback given different (more physical) hydrodynamical models, \eg FIRE \citep{Hopkins:2017ycn}.

Ref.~\cite{Fernandez:2023grg} re-analyzed the eBOSS flux power spectrum with simulations including spatial fluctuations in helium reionization \citep{Bird:2023evb}, finding a higher value of the power spectrum tilt that would disfavor running models relative to ULAs; although, their astrophysical model is not completely consistent between flux power spectrum and high-resolution spectrum IGM temperature \citep{Gaikwad:2020eip} measurements\footnote{This inconsistency may be relieved by varying cosmology when fitting the high-resolution data [Simeon Bird, private communication].}. Ref.~\cite{2024arXiv241205372W} also re-analyzed the data, finding a higher value of the tilt, albeit with a strong sensitivity in their inference to the prior distribution on the mean transmitted flux. Ref.~\cite{Goldstein:2023gnw} performed \(\Lambda\)CDM and early dark energy analyses using the compressed eBOSS likelihood; we use the compression as tabulated by Ref.~\cite{Goldstein:2023gnw} and therefore find our \(\Lambda\)CDM constraints are the same. For upcoming Dark Energy Spectroscopic Instrument (DESI) data \citep{DESI:2023pir}, where the statistical precision will reach percent level, it is crucial that the fidelity of IGM simulations further increases accounting for scale-dependent effects like HCDs. Coupled with higher-accuracy IGM simulations, it will be feasible to distinguish further between the models we have analyzed and astrophysical uncertainties like HCDs. LSS measurements of the matter power spectrum, \eg the Rubin Observatory \citep{LSSTDarkEnergyScience:2012kar} and Simons Observatory \citep{SimonsObservatory:2018koc}, will be highly complementary.

Models that reduce the eBOSS Lyman-alpha forest tension suppress power at higher wavenumbers \(k \sim 10\,h\,\mathrm{Mpc}^{-1}\) that are probed by, \eg higher-resolution Lyman-alpha forest \citep{Archidiacono:2019wdp,Boera:2018vzq,Day:2019joh,Karacayli:2021jeg,Irsic:2017sop,Rogers:2020ltq,Rogers:2021byl,Rogers:2020cup} and Milky Way sub-structure \citep{DES:2020fxi,Enzi:2020ieg,Banik:2019smi}. We anticipate that the \(\sim 10 \%\) suppression implied by eBOSS is consistent at the current sensitivity of higher-resolution flux power spectrum measurements from \textit{Keck} and VLT (which have \(\sim 25 \%\) uncertainty); the XQ-100 Lyman-alpha forest dataset \citep{Irsic:2017sop}, which probes smaller scales than eBOSS, infers an even lower value of \(n_\mathrm{lin}\), which may favor power spectrum running models\footnote{We do not use the compressed XQ-100 likelihood \citep{Goldstein:2023gnw} as it has been compressed at the eBOSS pivot wavenumber despite XQ-100 data being sensitive to smaller scales. This approach necessarily means that a power-law matter power spectrum model has been assumed from eBOSS to XQ-100 scales, which does not apply to the extended cosmological models that we consider. We defer to future work a detailed analysis with dedicated simulations.}. Hints of new physics were found in the HIRES/MIKE quasar sample \cite{Hooper:2021rjc,Hooper:2022byl}. Ref.~\cite{Esposito:2022plo} identified a \(\sim 3.3 \sigma\) discrepancy in the inference of \(\sigma_8\) between XQ-100 and HIRES/MIKE Lyman-alpha forest and galaxy cluster number counts from the South Pole Telescope Sunyaev-Zel'dovich effect \citep{Bleem_2015}; they discuss HCD contamination in the context of this tension. We defer a detailed joint analysis of eBOSS and small-scale structure probes to future work.

Finally, we find in our joint analyses that no model recovers low enough \(\sigma_8\) to restore consistency with the \(\Lambda\)CDM values inferred from galaxy weak lensing (see Fig.~\ref{fig:contours_tension} and the Supplemental Material for more details). However, the same argument for the Lyman-alpha forest applies to weak lensing; we must assess consistency at the scales and redshifts to which weak lensing probes are sensitive. Our work demonstrates that it is critical to use model parameterizations that capture the information content of a given experiment; we present a concrete step in this direction.

\textbf{Acknowledgments} -- The authors thank Simeon Bird, Andreu Font-Ribera, Julien Lesgourgues, Riccardo Murgia and Nils Schöneberg for valuable comments. The authors also thank Cosmology Talks mini-workshops where elements of this project were conceived. KKR is supported by an Ernest Rutherford Fellowship from the UKRI Science and Technology Facilities Council (grant number ST/Z510191/1). The Dunlap Institute is funded through an endowment established by the David Dunlap family and the University of Toronto. VP is supported by funding from the European Research Council (ERC) under the European Union’s HORIZON-ERC-2022 (grant agreement no.~101076865). VP is also supported by the European Union’s Horizon 2020 research and innovation program under the Marie Sk{\l}odowska-Curie grant agreement no.~860881-HIDDeN. The authors acknowledge the use of computational resources from the Excellence Initiative of Aix-Marseille University (A*MIDEX) of the Investissements d’Avenir program and from the LUPM's cloud computing infrastructure founded by Ocevu labex and France-Grilles.

\subsection{\label{sec:supplement}Supplemental Material}

\textbf{eBOSS Lyman-alpha forest flux power spectrum likelihood} -- Here, we give details on the astrophysical model that is marginalized in the compressed eBOSS Lyman-alpha forest flux power spectrum likelihood that we use. The compressed likelihood was obtained by the eBOSS Collaboration using a set of 28 hydrodynamical simulations \citep{Borde:2014xsa} varying cosmological and IGM parameters, each a splicing of three simulations at different resolutions with a maximum box size of \((100 h^{-1}\,\mathrm{Mpc})^3\) in order to span efficiently the full range of wavenumbers, all run using the smoothed particle hydrodynamics code \texttt{GADGET-3} \citep{2005MNRAS.364.1105S,Springel:2000yr}.

Cosmological parameters (\(\sigma_8, n_\mathrm{s}, \Omega_\mathrm{m}, H_0\)) and IGM parameters (the temperature \(T_0\) at mean density at \(z = 3\), the spectral index \(\gamma\) at \(z = 3\) that defines the relation between temperature \(T\) and baryon overdensity \(\delta\): \(T = T_0 \delta^{\gamma - 1}\)) are varied across the simulations. \(T_0\) and \(\gamma\) are allowed to vary with \(z\) such that \(T_0(z) = T_0 \left(\frac{1 + z}{4}\right)^{\eta_{T_0}}\), where \(\eta_{T_0}\) has different values for \(z > 3\) and \(z < 3\), and \(\gamma(z) = \gamma \left(\frac{1 + z}{4}\right)^{\eta_\gamma}\). Variations in the photo-ionization rate are captured by post-processing the effective optical depth (\(\tau_\mathrm{eff} \equiv - \mathrm{ln} \langle F \rangle\), where \(\langle F \rangle\) is the mean flux transmission) such that \(\tau_\mathrm{eff} = A_\tau (1 + z)^{\eta_\tau}\). Further nuisance parameters are varied in generating mock flux power spectra: two amplitudes for the correlated absorption of Ly-\(\alpha\) with SiII and SiIII, a noise term, the spectrograph resolution, bias from the splicing technique, ultraviolet background fluctuations, residual contamination from high-column-density (HCD) absorbers, supernovae and active galactic nuclei feedback.

\begin{figure}
\includegraphics[width=\columnwidth]{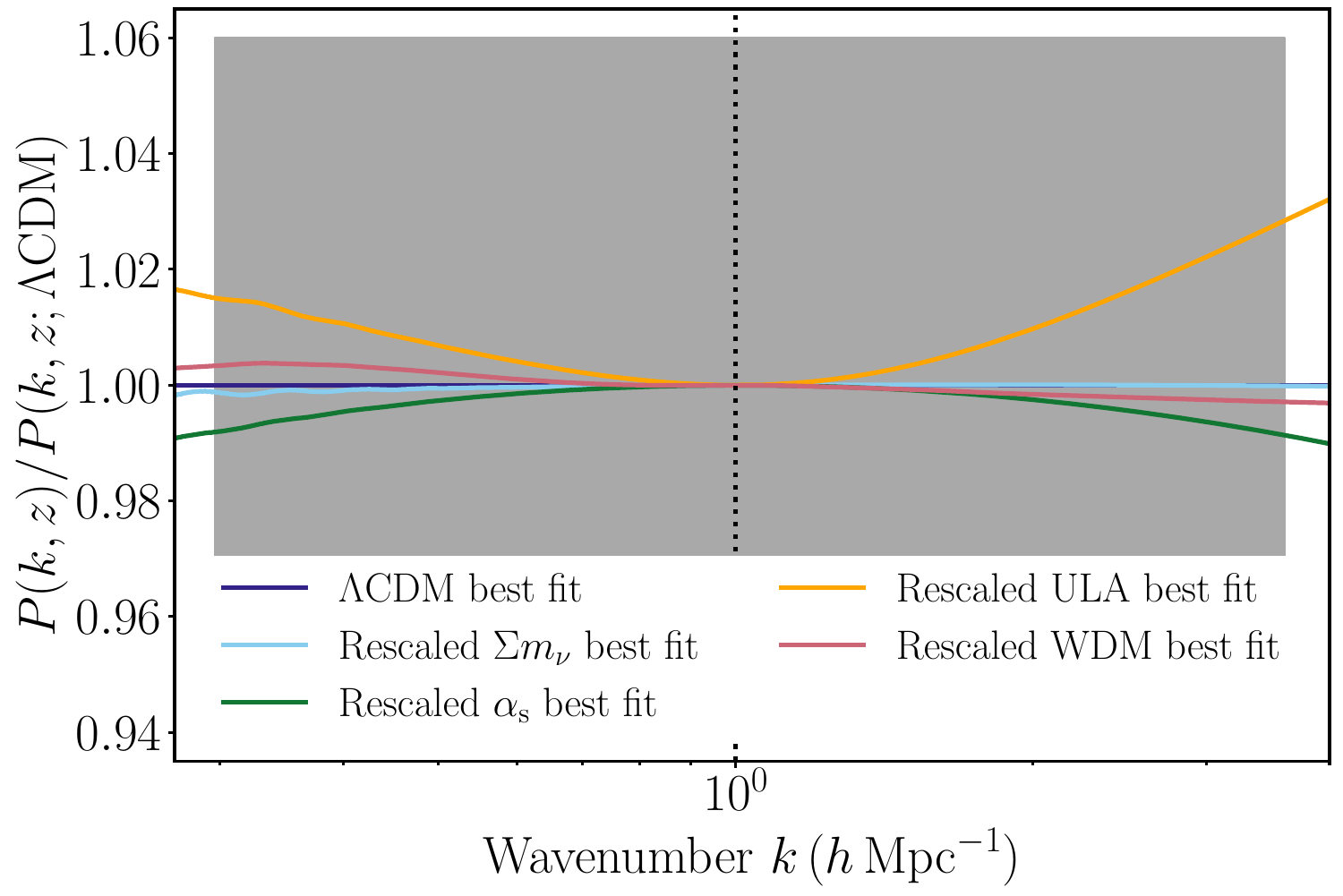}
\caption{\label{fig:model_residuals}Best fit linear matter power spectra at \(z = 3\) rescaled to have the same values of \(\Delta_\mathrm{lin}^2, n_\mathrm{lin}\) and shown in ratio to the \(\Lambda\)CDM model. After rescaling, all best fit models are indistinguishable within the \(1 \sigma\) sensitivity of the eBOSS Lyman-alpha forest flux power spectrum (approximately indicated by the gray area). This consistency demonstrates that all the cosmological information in the eBOSS data is well captured by \(\Delta_\mathrm{lin}^2, n_\mathrm{lin}\) for the viable model parameter spaces that we consider.}
\end{figure}
Figure \ref{fig:model_residuals} demonstrates that the compressed eBOSS Lyman-alpha forest likelihood is accurate for the cosmological models that we consider given the precision of eBOSS data. We take the best-fit linear matter power spectra for each model and rescale them such that they have the same values of \(\Delta_\mathrm{lin}^2, n_\mathrm{lin}\) as the best-fit \(\Lambda\)CDM model. Once the information in these two parameters is accounted, there is negligible difference between the model power spectra compared to the \(\sim 6 \%\) precision of eBOSS data (conservatively estimated from the constrained uncertainty on \(\Delta_\mathrm{lin}^2\)). The range of wavenumbers to which eBOSS is sensitive is estimated from the velocity wavenumbers measured in the flux power spectrum, although the flux power spectrum is theoretically sensitive to all wavenumbers since it is an integral over modes transverse to the line of sight. The range is calculated by Ref.~\cite{Chabanier:2019eai} which demonstrates that the sensitivity of the eBOSS 1D flux power spectrum to the linear matter power spectrum is negligible for smaller or larger wavenumbers given the poor observational sampling of modes transverse to the line of sight. To be additionally conservative, we confirm that our extended models remain sufficiently close to a power law over an even longer range of wavenumbers \(0.1\,h\,\mathrm{Mpc}^{-1} < k < 8\,h\,\mathrm{Mpc}^{-1}\). The validity of the compression is further confirmed by explicitly demonstrating that the posterior distribution is identical whether inferred using the full flux power spectrum or the compressed likelihood \citep{Pedersen:2022anu}.

The compression is invalid for models with sharp features in the matter power spectrum, \eg ULA or WDM models where the new particle forms the entirety of the dark matter. We deliberately restrict ourselves to mixed dark matter models, with a prior set to preclude these features forming. Given the anticipated precision of upcoming DESI data, it may be necessary to add a parameter to the compression capturing curvature in the power spectrum, in particular to account for mixed ULA models. This extra information may also distinguish further between cosmological models.

\begin{figure*}
\includegraphics[width=2\columnwidth]{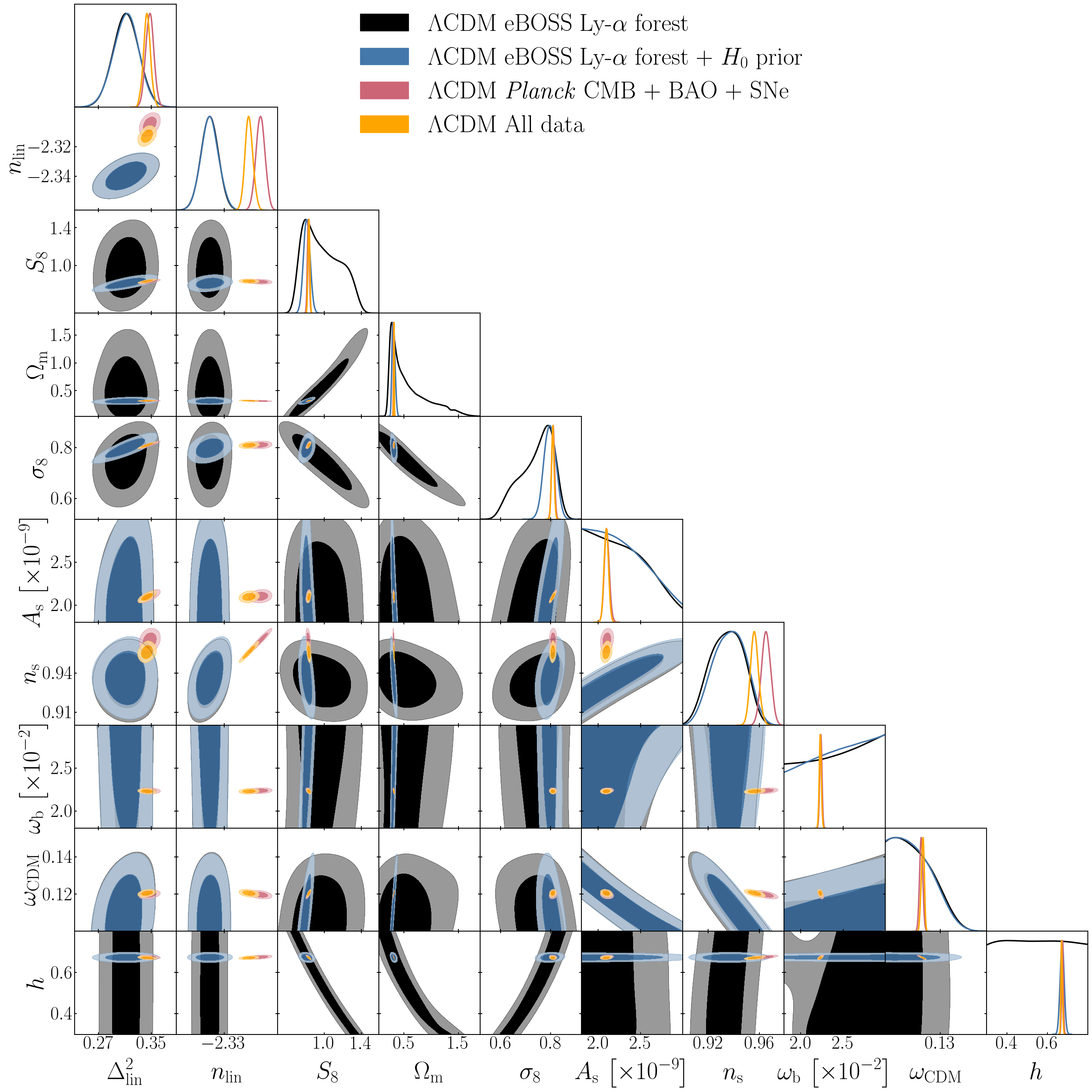}
\caption{\label{fig:triangle}1D and 2D marginalized posterior distributions of cosmological parameters given the \(\Lambda\)CDM model and the indicated data combinations and prior distributions. ``All data'' refers to the combination of \textit{Planck} CMB + BAO + SNe + eBOSS Lyman-alpha forest. The darker and lighter shaded areas respectively indicate the 68\% and 95\% credible regions. The dimensionless Hubble parameter \(h = \frac{H_0}{100\,\mathrm{km}\,\mathrm{s}^{-1}\,\mathrm{Mpc}^{-1}}\).}
\end{figure*}
\begin{figure*}

\includegraphics[width=2\columnwidth]{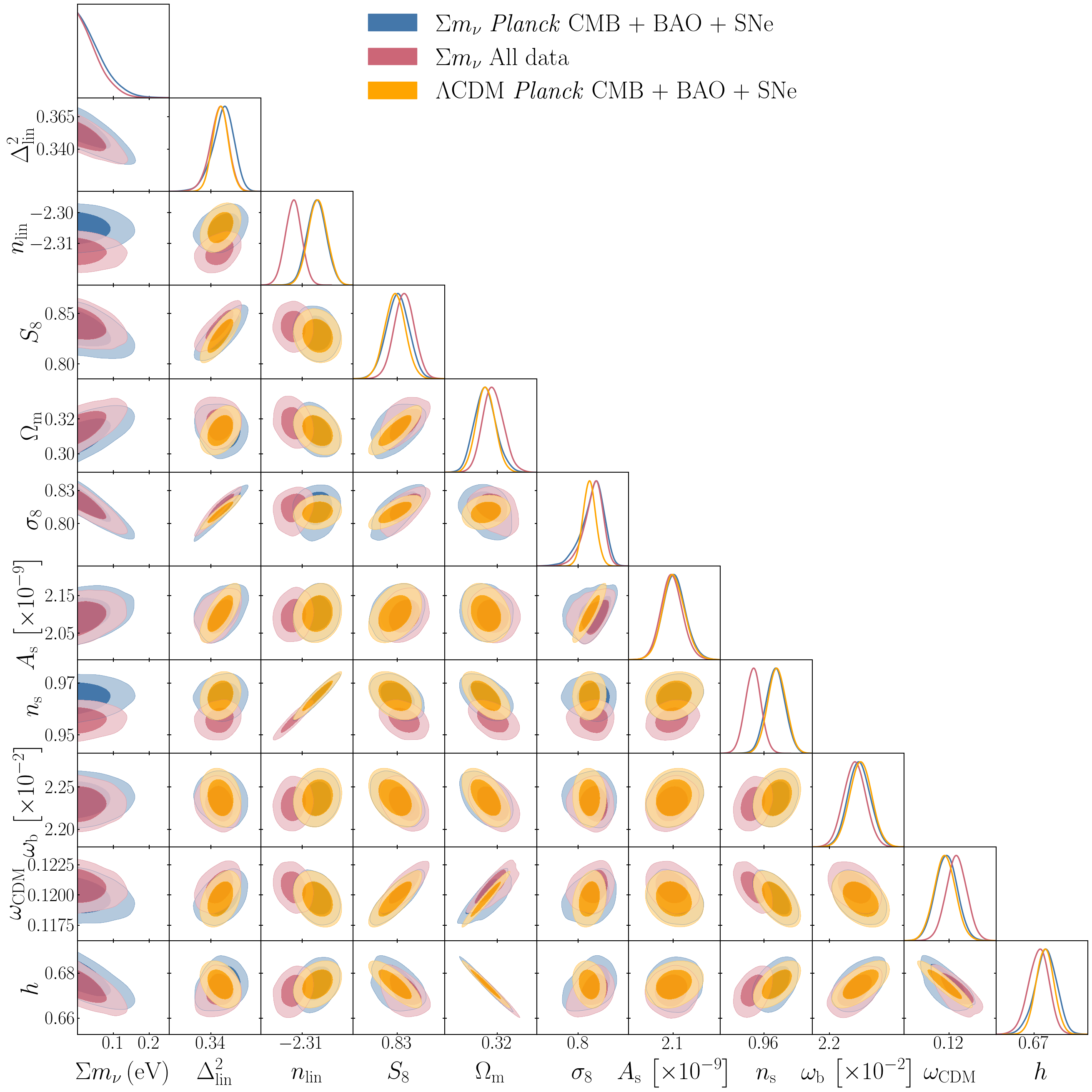}
\caption{\label{fig:triangle2}As Fig.~\ref{fig:triangle} except comparing the \(\Sigma m_\nu\) and \(\Lambda\)CDM models.}
\end{figure*}

\begin{table*}[]
    \centering
    \begin{tabular}{l|c|c}
    \hline
    \multicolumn{3}{c}{$\Sigma m_\nu$}\\
     \hline
& \textit{Planck} CMB + BAO + SNe & + eBOSS Lyman-alpha forest
\\     \hline

	$\Sigma m_\nu\,(\mathrm{eV})$ & $< 0.129 (0.001)$
	 &  $< 0.110 (0.002)$
	 \\
\hline
$h$
	 & $0.6753(0.6785)^{+0.0052}_{-0.0045}$ 
	 & $0.6721(0.6747)^{+0.0049}_{-0.0043}$ 
	 \\
$\omega_{\rm CDM}$
	 & $0.11980(0.12005)\pm 0.00092$ 
	 & $0.12064(0.12087)\pm 0.00087$ 
	 \\
$10^{2}\omega_\mathrm{b}$
	 & $2.236(2.236)\pm 0.013$ 
	 & $2.230(2.229)\pm 0.013$ 
	 \\
$10^{9}A_\mathrm{s}$
	 & $2.104(2.099)\pm 0.031$ 
	 & $2.097(2.087)\pm 0.029$ 
	 \\
$n_\mathrm{s}$
	 & $0.9646(0.9653)\pm 0.0037$ 
	 & $0.9558(0.9555)\pm 0.0031$ 
	 \\
$\tau$
	 & $0.0554(0.0542)\pm 0.0073$ 
	 & $0.0523(0.05)\pm 0.0070$ 
	 \\
\hline
$\Delta_\mathrm{lin}^2$
	 & $0.3495(0.3578)^{+0.0095}_{-0.0071}$ 
	 & $0.3465(0.3515)^{+0.0078}_{-0.0065}$ 
	 \\
$n_\mathrm{lin}$
	 & $-2.3052(-2.3037)\pm 0.0032$ 
	 & $-2.3129(-2.3126)\pm 0.0027$ 
	 \\
$S_8$
	 & $0.831(0.836)\pm 0.011$ 
	 & $0.837(0.841)\pm 0.010$ 
	 \\
$\Omega_\mathrm{m}$
	 & $0.3131(0.3094)^{+0.0059}_{-0.0067}$ 
	 & $0.3176(0.3146)^{+0.0057}_{-0.0064}$ 
	 \\

  \hline
    \end{tabular}
    \caption{1D marginalized posterior mean, maximum likelihood (in brackets) and 68\% credible limits (or 95\% upper limits) on cosmological parameters for the \(\Sigma m_\nu\) model given \textit{Planck} CMB + BAO + SNe data without (\textit{left}) and with (\textit{right}) the addition of eBOSS Lyman-alpha forest.}
    \label{tab:mnu}
\end{table*}

\begin{figure*}
\includegraphics[width=2\columnwidth]{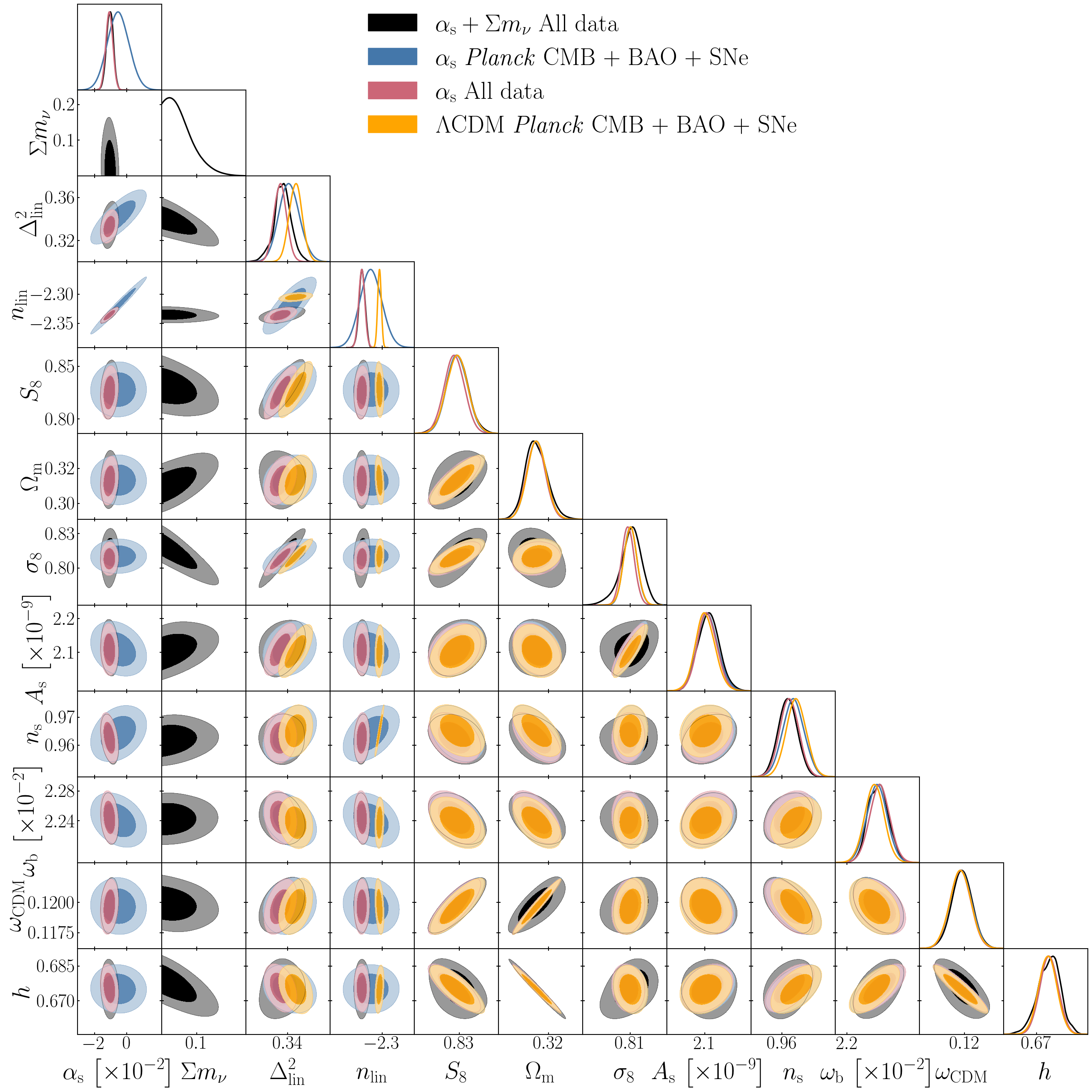}
\caption{\label{fig:triangle3}As Fig.~\ref{fig:triangle} except comparing the \(\alpha_\mathrm{s}\) and \(\Lambda\)CDM models. \(\Sigma m_\nu\) is given in units of eV.}
\end{figure*}

\begin{table*}[]
    \centering
    \begin{tabular}{l|c|c}
    \hline
    \multicolumn{3}{c}{$\alpha_\mathrm{s}$}\\
     \hline
& \textit{Planck} CMB + BAO + SNe & + eBOSS Lyman-alpha forest
\\     \hline
 $\alpha_\mathrm{s}$
	 & $-0.0053(-0.0041)\pm 0.0066$ 
	 & $-0.0108(-0.0106)\pm 0.0022$ 
	 \\
\hline

$h$
	 & $0.6752(0.6753)\pm 0.0040$ 
	 & $0.6755(0.6754)\pm 0.0040$ 
	 \\
$\omega_{\rm CDM}$
	 & $0.11969(0.11966)\pm 0.00091$ 
	 & $0.11966(0.11974)\pm 0.00088$ 
	 \\
$10^{2}\omega_\mathrm{b}$
	 & $2.241(2.242)\pm 0.014$ 
	 & $2.245(2.245)\pm 0.014$ 
	 \\
$10^{9}A_\mathrm{s}$
	 & $2.109(2.108)\pm 0.031$ 
	 & $2.108(2.109)\pm 0.030$ 
	 \\
$n_\mathrm{s}$
	 & $0.9642(0.965)\pm 0.0039$ 
	 & $0.9626(0.9626)\pm 0.0035$ 
	 \\
$\tau$
	 & $0.0563(0.0563)^{+0.0068}_{-0.0077}$ 
	 & $0.0556(0.0557)\pm 0.0072$
  \\
\hline

$\Delta_\mathrm{lin}^2$
	 & $0.3416(0.3433)\pm 0.0097$ 
	 & $0.3332(0.3339)\pm 0.0060$ 
	 \\
$n_\mathrm{lin}$
	 & $-2.320(-2.316)\pm 0.019$ 
	 & $-2.3358(-2.3351)\pm 0.0054$ 
	 \\
$S_8$
	 & $0.828(0.827)\pm 0.010$ 
	 & $0.8254(0.8263)\pm 0.0098$ 
	 \\
$\Omega_\mathrm{m}$
	 & $0.3133(0.3130)\pm 0.0055$ 
	 & $0.3130(0.3132)\pm 0.0054$ 
	 \\
  \hline
    \end{tabular}
    \caption{As Table \ref{tab:mnu} except for the \(\alpha_\mathrm{s}\) model.}
    \label{tab:alpha_s}
\end{table*}

\begin{figure*}
\includegraphics[width=2\columnwidth]{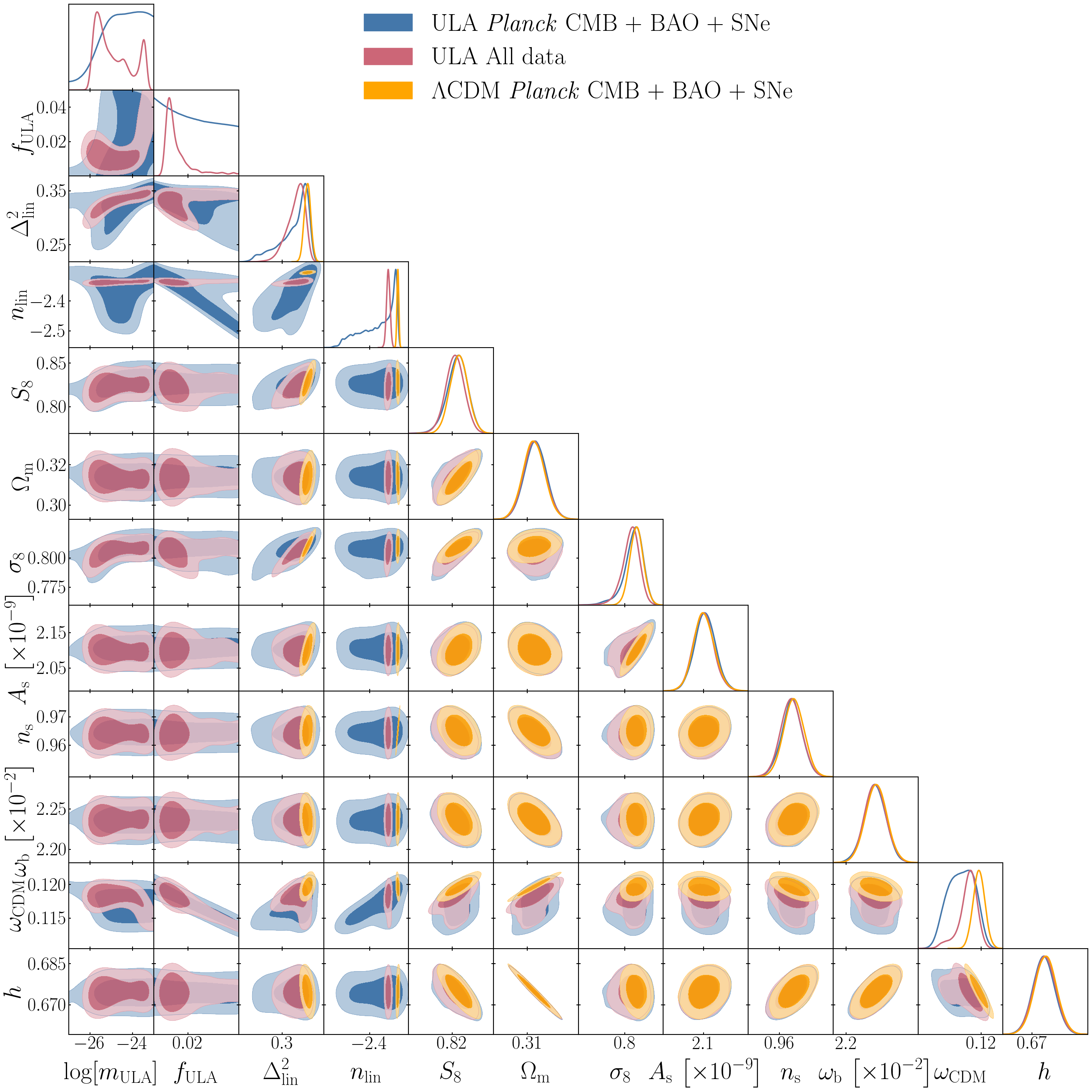}
\caption{\label{fig:triangle4}As Fig.~\ref{fig:triangle} except comparing the ULA and \(\Lambda\)CDM models. \(m_\mathrm{ULA}\) is given in units of eV.}
\end{figure*}

\begin{table*}[]
    \centering
    \begin{tabular}{l|c|c}
    \hline
    \multicolumn{3}{c}{ULA}\\
     \hline
& \textit{Planck} CMB + BAO + SNe & + eBOSS Lyman-alpha forest
\\     \hline
$\mathrm{log} [m_\mathrm{ULA}\,({\rm eV})]$
	 & Unconstrained (-23.2)
	 & $-24.9(-25.6)^{+1.5}_{-1.1}$ 
	 \\
$f_\mathrm{ULA}$
	 & Unconstrained (0.005)
	 & $0.0146(0.0127)^{+0.0014}_{-0.0086}$ 
	 \\
  \hline

$h$
	 & $0.6744(0.6748)\pm 0.0040$ 
	 & $0.6745(0.6747)\pm 0.0041$ 
	 \\
$\Omega_\mathrm{CDM}$
	 & $0.2574(0.2616)\pm 0.0062$ 
	 & $0.2593(0.2596)\pm 0.0056$ 
	 \\
$10^{2}\omega_\mathrm{b}$
	 & $2.236(0.022)\pm 0.013$ 
	 & $2.237(0.022)\pm 0.014$ 
	 \\
$10^9A_\mathrm{s}$
	 & $2.105(2.105)\pm 0.030$ 
	 & $2.101(2.104)\pm 0.031$ 
	 \\
$n_\mathrm{s}$
	 & $0.9644(0.9656)\pm 0.0037$ 
	 & $0.9643(0.9651)\pm 0.0039$ 
	 \\
$\tau$
	 & $0.0558(0.0555)\pm 0.0071$ 
	 & $0.0549(0.0555)\pm 0.0076$ 
	 \\
  \hline
$\Delta_\mathrm{lin}^2$
	 & $0.323(0.349)^{+0.032}_{-0.011}$ 
	 & $0.325(0.314)^{+0.020}_{-0.011}$ 
	 \\
$n_\mathrm{lin}$
	 & $-2.370(-2.306)^{+0.068}_{-0.045}$ 
	 & $-2.3374(-2.3396)\pm 0.0060$ 
	 \\
$S_8$
	 & $0.826(0.83)^{+0.013}_{-0.010}$ 
	 & $0.823(0.822)\pm 0.011$ 
	 \\
$\Omega_\mathrm{m}$
	 & $0.3140(0.3136)\pm 0.0055$ 
	 & $0.3137(0.3136)\pm 0.0056$ 
	 \\

\hline 
\end{tabular}
    \caption{As Table \ref{tab:mnu} except for the ULA model.}
    \label{tab:ULA}
\end{table*}

\begin{figure*}
\includegraphics[width=2\columnwidth]{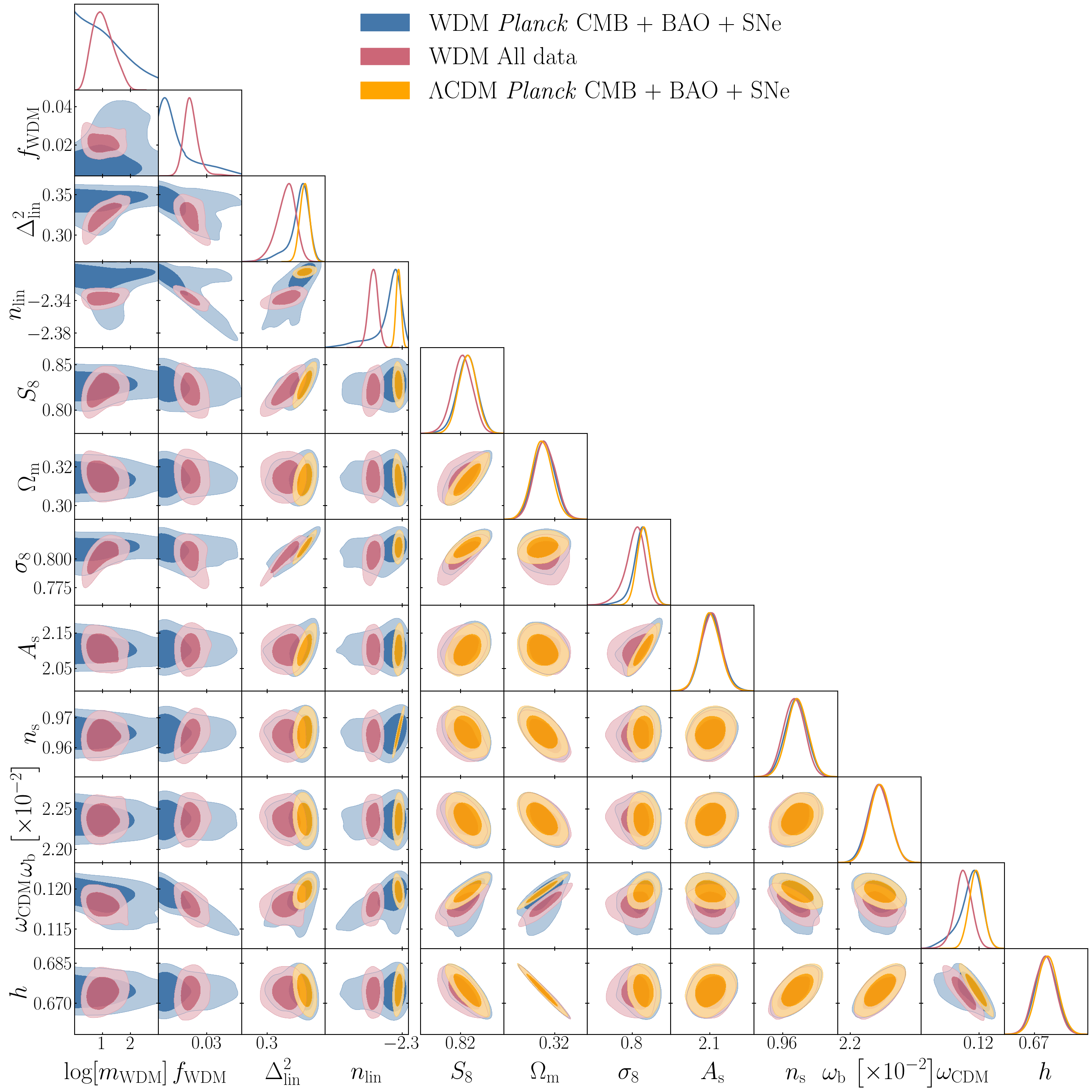}
\caption{\label{fig:triangle5}As Fig.~\ref{fig:triangle} except comparing the WDM and \(\Lambda\)CDM models. \(m_\mathrm{WDM}\) is given in units of eV and \(A_\mathrm{s}\) in units of \(10^{-9}\).}
\end{figure*}

\begin{table*}[]
    \centering
    \begin{tabular}{l|c|c}
    \hline
    \multicolumn{3}{c}{WDM}\\
     \hline
& \textit{Planck} CMB + BAO + SNe & + eBOSS Lyman-alpha forest
\\     \hline
	$\log[m_\mathrm{WDM}\,({\rm eV})]$ & Unconstrained (0.21) 
	 &  $1.01(0.95)^{+0.30}_{-0.44}$
	 \\
$\frac{T_\mathrm{WDM}}{T_\mathrm{CMB}}$
	 & $< 0.261 (0.001)$ 
	 &   $0.194(0.194)^{+0.043}_{-0.074}$
	 \\
$f_{\rm WDM}$
	 & $<0.049 (0.005)$
	 &  $0.0219(0.0219)^{+0.0030}_{-0.0042}$
	 \\
\hline
$h$
	 & $0.6745(0.6745)\pm 0.0041$ 
	 & $0.6741(0.6748)\pm 0.0040$ 
	 \\
$\omega_\mathrm{CDM}$
	 & $0.1189(0.1198)^{+0.0018}_{-0.0009}$ 
	 & $0.1180(0.1178)\pm0.0011$ 
	 \\
$10^{2}\omega_\mathrm{b}$
	 & $2.237(2.237)\pm 0.014$ 
	 & $2.237(2.237)\pm 0.013$ 
	 \\
$10^9A_\mathrm{s}$
	 & $2.105(2.104)\pm 0.030$ 
	 & $2.105(2.104)\pm 0.029$ 
	 \\
$n_\mathrm{s}$
	 & $0.9648(0.9657)\pm 0.0037$ 
	 & $0.9640(0.9655)\pm 0.0036$ 
	 \\
$\tau$
	 & $0.0558(0.0555)\pm0.0072$ 
	 & $0.0558(0.0554)\pm 0.0068$ 
	 \\
  \hline
  $\Delta_\mathrm{lin}^2$
	 & $0.341(0.349)^{+0.014}_{-0.005}$ 
	 & $0.324(0.322)^{+0.013}_{-0.010}$ 
	 \\
$n_\mathrm{lin}$
	 & $-2.316(-2.304)^{+0.015}_{-0.0009}$ 
	 & $-2.3367(-2.3382)\pm 0.0058$ 
	 \\
$S_8$
	 & 	$0.827(0.830)\pm 0.011$
	 & 	$0.822(0.821)\pm 0.012$
	 \\
$\Omega_\mathrm{m}$
	 & $0.3141(0.3139)\pm 0.0055$ 
	 & $0.3148(0.3136)\pm 0.0055$ 
	 \\

\hline 

\end{tabular}
    \caption{As Table \ref{tab:mnu} except for the WDM model.}
    \label{tab:WDM}
\end{table*}

\begin{table*}[]
\begin{ruledtabular}
    \centering
    \begin{tabular}{lccc c c c c c c c}
        Data & Model & Tot. $\chi^2$ & TTTEEE & Low-$\ell$ EE & Low-$\ell$ TT & \(\phi \phi\) & SNe & 6dF/MGS & BOSS & Ly$\alpha$ \\
        \hline \rule{0pt}{3ex}
        \multirow{5}{*}{\textit{Planck} CMB + BAO + SNe} & $\Lambda$CDM & 4192.15& 2346.62 &  396.31& 23.32& 8.76 &1410.95 & 1.06 & 5.12 & $-$ \\
      & $\Sigma m_\nu$   & 4191.75 & 2345.41 & 396.00 & 23.42 & 8.88  & 1411.55 & 1.29& 4.19 & $-$  \\
             & \(\alpha_\mathrm{s}\) & 4191.90 &  2347.16&  396.33 & 22.41 & 8.83 & 1410.93 & 1.05 & 5.19 & $-$ \\
              & ULA  & 4192.18 &  2346.73 & 396.25 & 23.30 & 8.75 & 1410.96 & 1.06 & 5.13 &   $-$ \\
              & WDM  & 4192.11 & 2346.66 & 396.25 & 23.29 & 8.75 &  1410.93 & 1.05 & 5.19 & $-$ \\
        \hline
        \multirow{5}{*}
        {+ eBOSS Lyman-alpha forest} & $\Lambda$CDM & 4220.23& 2351.02 &   395.99& 25.72& 8.58 & 1410.38 &0.85 & 6.86 & 20.81 \\
      & $\Sigma m_\nu$   & 4218.33 &2349.77 & 395.69 & 25.79 & 8.87  & 1410.86 & 1.02 & 5.42 & 20.92 \\
             & \(\alpha_\mathrm{s}\) & 4194.62 & 2349.24 & 396.30 &  21.28 & 9.0 & 1410.93 & 1.04 & 5.25 & 1.58 \\
              & ULA  &  4192.65 & 2346.99 & 396.26 & 23.40 & 8.80 & 1410.96 & 1.06 & 5.13 & 0.05  \\
              & WDM  &  4193.15 & 2347.07 & 396.25  & 23.31 & 8.84 & 1410.98 & 1.07 & 5.08 & 0.39  \\
    \end{tabular}
    \end{ruledtabular}
    \caption{Minimum \(\chi^2\) given the indicated data combination and cosmological model broken down by data component. Ly\(\alpha\) refers to the eBOSS Lyman-alpha forest likelihood.}
    \label{tab:chi2}
\end{table*}

\textbf{Other likelihoods and analysis details} -- For CMB data, we use the \texttt{Plik} low-multipole $\ell$ TT and EE, high-$\ell$ TTTEEE and conservative lensing \(\phi \phi\) likelihoods from the {\it Planck}~2018 data release \citep{Planck:2018vyg,Planck:2019nip}. We use BAO measurements from the CMASS and LOWZ galaxy samples of BOSS Data Release 12 at $z = 0.38$, $0.51$ and $0.61$ \citep{BOSS:2016wmc}, the 6dF Galaxy Survey (6dFGS) at $z = 0.106$ \citep{Beutler:2011hx} and SDSS Data Release 7 Main Galaxy Sample (SDSS DR7 MGS) at $z = 0.15$ \citep{Ross:2014qpa}; and the Pantheon+ catalog of uncalibrated SNe luminosity distances \citep{Brout:2022vxf}. This compendium of CMB, BAO and SNe data constitutes a larger-scale complement to the smaller scales probed by the eBOSS flux power spectrum. We vary nuisance foreground parameters in the \textit{Planck} likelihood and SNe calibrations in the Pantheon+ likelihood.

We use a uniform prior distribution on baseline \(\Lambda\)CDM parameters: the Hubble constant $H_0$\footnote{For technical reasons, we sometimes instead use the angular size of the sound horizon $\theta_\mathrm{s}$.}, the physical baryon $\omega_\mathrm{b}$ and CDM $\omega_{\rm CDM}$ energy densities, the amplitude $A_\mathrm{s}$ and tilt $n_\mathrm{s}$ of the primordial power spectrum and the reionization optical depth $\tau$. For the extended cosmological parameters, we set uniform prior ranges $\mathrm{log} [m_{\rm ULA}\,(\mathrm{eV})] \in [-27,-23], f_{\rm ULA} \in [0,0.05], \mathrm{log} [m_{\rm WDM}\,(\mathrm{eV})] \in [0, 3], \frac{T_{\rm WDM}}{T_{\rm CMB}} \in [0,1]$, where \(T_\mathrm{CMB}\) is the CMB temperature, with the additional constraint $f_{\rm WDM} < 0.05$. Unless specified otherwise, we model free-streaming neutrinos as two massless and one massive neutrino with the sum of neutrino masses $\Sigma m_\nu=0.06$ eV.

We use the cosmological inference package \texttt{MontePython-v3}\footnote{\url{https://github.com/brinckmann/montepython_public}.} \citep{Audren:2012wb,Brinckmann:2018cvx} interfaced with a modified version\footnote{\url{https://github.com/PoulinV/AxiCLASS}.} of the Einstein-Boltzmann solver \texttt{CLASS}\footnote{\url{https://lesgourg.github.io/class_public/class.html}.} \cite{Lesgourgues:2011re,Blas:2011rf} in order to perform a Markov chain Monte Carlo (MCMC) sampling of the posterior distribution using a Metropolis Hasting algorithm. We model ULAs following the effective fluid approach from the modified Einstein-Bolztmann solver \texttt{AxionCAMB} \citep{Hlozek:2014lca} and implemented in \texttt{AxiCLASS}\footnote{An improved fluid model was derived in Ref.~\cite{Passaglia:2022bcr} but the difference is negligible given the precision of our data.} \citep{Poulin:2018dzj,Smith:2019ihp}. We consider MCMC chains converged when the Gelman-Rubin statistic $|R - 1|\lesssim 0.05$ \citep{Gelman:1992zz}. To analyze MCMC chains and visualize posteriors, we use the \texttt{GetDist} package \cite{Lewis:2019xzd}. We find minimum $\chi^2$ values using the simulated annealing method described in the appendix of Ref.~\cite{Schoneberg:2021qvd}.

\textbf{Full sets of marginalized posteriors} -- Figures \ref{fig:triangle} to \ref{fig:triangle5} show full sets of 1D and 2D marginalized posteriors of cosmological parameters given the different cosmological models and data combinations. Fig.~\ref{fig:triangle} shows the case for the \(\Lambda\)CDM model. The two-parameter compressed eBOSS Lyman-alpha forest-only likelihood is, unsurprisingly, unable to constrain the six-parameter baseline cosmology. Even after adding information through a Gaussian prior on \(H_0\) inspired by \textit{Planck} CMB results (\(H_0 \sim \mathcal{N}(67.3, 1)\,\mathrm{km}\,\mathrm{s}^{-1}\,\mathrm{Mpc}^{-1}\)) as done by the eBOSS Collaboration, the other standard \(\Lambda\)CDM parameters remain poorly constrained. This is different from the results presented in Ref.~\cite{Palanque-Delabrouille:2019iyz}: after additionally fixing the baryon energy density, they simultaneously constrain \(\sigma_8\), \(n_\mathrm{s}\) and the total matter energy density \(\Omega_\mathrm{m}\) using a full flux power spectrum likelihood.

However, this contradicts the results of Refs.~\cite{Pedersen:2020kaw,Pedersen:2019ieb,Pedersen:2022anu,McDonald_2000} which demonstrate that full flux power spectrum and compressed likelihoods return identical posteriors on \(\Lambda\)CDM parameters -- and the re-analysis of eBOSS flux power spectra \citep{Fernandez:2023grg} that constrains two cosmological parameters (the amplitude and tilt of the primordial power spectrum defined at \(k = 0.78 h\,\mathrm{Mpc}^{-1}\)) after using a uniform prior on \(\Omega_\mathrm{m} h^2\) restricting it to values compatible with \textit{Planck} and fixing \(\omega_\mathrm{b}\)\footnote{Ref.~\cite{Fernandez:2023grg} constrains \(H_0\) as an artifact of converting from velocity to spatial units in a finite simulation box, \ie there is no physical information on \(H_0\) in the data.}. In testing, we find that the eBOSS-only \(\Lambda\)CDM constraints are highly sensitive to the width of uniform priors on cosmological parameters. Indeed, we recover the same degeneracy as the eBOSS Collaboration between \(n_\mathrm{s}\) and \(\Omega_\mathrm{m}\) and the same mild discrepancy with \textit{Planck} contours in this plane, except that the constraint along the degeneracy direction is prior-dependent. The full flux power spectrum likelihood used by the eBOSS Collaboration is not publicly available and so we defer to future work a detailed comparison of eBOSS-only constraints, in particular accounting for the effect of prior distributions.

Our measurement of tension in the \(\Delta_\mathrm{lin}^2, n_\mathrm{lin}\) parameters is in exact agreement with results of the eBOSS Collaboration as we use the compressed likelihood as calculated by them. Further, we reproduce the combined \textit{Planck}, BAO and eBOSS constraints in the \(\Lambda\)CDM and varying neutrino mass models (Fig.~\ref{fig:triangle2}) of Ref.~\cite{Palanque-Delabrouille:2019iyz}. This result arises since, once data are sufficiently constraining on the full six- or seven-parameter models, the posterior loses sensitivity to the choice of prior. Therefore, our primary conclusions regarding the tension in the \(\Delta_\mathrm{lin}^2, n_\mathrm{lin}\) parameters and combined CMB, BAO, SNe and Lyman-alpha forest constraints on extended models is unaffected by the choice of prior or the use of a full or compressed likelihood.

Figure \ref{fig:triangle2} shows how increasing the sum of neutrino masses lowers both \(\Delta_\mathrm{lin}^2\) and \(\sigma_8\), but doesn't change \(n_\mathrm{lin}\) or \(n_\mathrm{s}\). When CMB + BAO + SNe data are added to the flux power spectrum, lower values of the spectral indices are only achieved by degrading the fit to the CMB + BAO + SNe dataset. In contrast, Fig.~\ref{fig:triangle3} demonstrates how reducing \(\alpha_\mathrm{s}\) lowers both \(\Delta_\mathrm{lin}^2\) and \(n_\mathrm{lin}\), while keeping \(\sigma_8\) and \(n_\mathrm{s}\) consistent with \textit{Planck} values. Therefore, when adding Lyman-alpha forest data, a preference for non-zero running is found. We also show in Fig.~\ref{fig:triangle3} the case where we vary both \(\Sigma m_\nu\) and \(\alpha_\mathrm{s}\) finding a weak degeneracy between the two parameters. The upper limit on $\Sigma m_\nu$ relaxes slightly from $\Sigma m_\nu < 0.110$ eV to $\Sigma m_\nu < 0.133$ eV once \(\alpha_\mathrm{s}\) is additionally varied. The degeneracies between ULA and \(\Lambda\)CDM parameters (Fig.~\ref{fig:triangle4}) are more non-trivial. There is a range of \(m_\mathrm{ULA}\) where \(\sigma_8\) is lowered (\(m_\mathrm{ULA} < 10^{-25}\,\mathrm{eV}\)) which has previously been found to alleviate the \(S_8\) tension \citep{Rogers:2023ezo}. There is equivalently a range of \(m_\mathrm{ULA}\) where \(\Delta_\mathrm{lin}^2\) and \(n_\mathrm{lin}\) are lowered (\(10^{-26}\,\mathrm{eV} < m_\mathrm{ULA} < 10^{-23}\,\mathrm{eV}\)). There is an equivalent effect for \(m_\mathrm{WDM} \lesssim 100\,\mathrm{eV}\) (Fig.~\ref{fig:triangle5}). Tables \ref{tab:mnu} to \ref{tab:WDM} give sets of marginalised constraints and best-fit values for cosmological parameters. Table \ref{tab:chi2} gives the minimum \(\chi^2\) for each data component. Only the running, ULA and WDM models give reasonable fits to the eBOSS Lyman-alpha forest likelihood.

\bibliography{eBOSS_letter}
\end{document}